\documentclass[5p]{elsarticle}
\usepackage{setspace}
\usepackage{amsmath}
\usepackage{amssymb}
\usepackage{color}
\usepackage{lineno,hyperref}
\modulolinenumbers[5]
\biboptions{sort&compress}

\journal{Extreme Mechanics Letters}

\newcommand{\lth}{\ensuremath{\ell_\text{th}}}
\newcommand{\lp}{\ensuremath{\ell_\text{p}}}
\newcommand{\dd}{\text{d}}
\newcommand{\ee}{\text{e}}
\newcommand{\ii}{\text{i}}

\newcommand{\Ybare}{\ensuremath{Y}}
\newcommand{\YT}{\ensuremath{Y_\text{R}}}
\newcommand{\kappabare}{\ensuremath{\kappa}}
\newcommand{\kappaT}{\ensuremath{\kappa_\text{R}}}
\newcommand{\kappasim}{\ensuremath{{\hat{\kappa}}}}

\begin{document}
\begin{frontmatter}

\title{Thermal buckling and symmetry breaking in thin ribbons under compression}

\author[1]{Paul~Z.~Hanakata\corref{cor1}}
\ead{paul.hanakata@gmail.com}
\author[2]{Sourav S. Bhabesh\fnref{fn1}}
\fntext[fn1]{Work completed prior to joining AWS.}
\author[3]{Mark J. Bowick}
\author[4]{David R. Nelson}
\author[5,6,3]{David Yllanes}

\address[1]{Department of Physics, Harvard University, Cambridge, MA 02138, USA}
\address[2]{Amazon Web Services (AWS), Washington DC Metro Area, USA}
\address[3]{Kavli Institute for Theoretical Physics, University of California, Santa Barbara, CA 93106, USA}
\address[4]{Department of Physics, Lyman Laboratory of Physics and School of Engineering and Applied Sciences,
Harvard University, Cambridge, MA 02138, USA}
\address[5]{Chan Zuckerberg Biohub, San Francisco, CA 94158, USA}
\address[6]{Instituto de Biocomputaci\'{o}n y F\'{\i}sica de Sistemas Complejos (BIFI), 50009 Zaragoza, Spain}
%\address[7]{Department of Physics, Syracuse University, Syracuse, NY, 13244, USA}
\cortext[cor1]{Corresponding author.}

%\address[1]{Department of Physics, Harvard University, Cambridge, MA 02138, USA}
%\address[2]{Department of Physics and Soft Matter Program, Syracuse University, Syracuse, NY, 13244, USA}
%\address[3]{Kavli Institute for Theoretical Physics, University of California, Santa Barbara, CA 93106, USA}
%\address[4]{Department of Physics, Lyman Laboratory of Physics and School of Engineering and Applied Sciences,
%Harvard University, Cambridge, MA 02138, USA}
%\address[5]{Chan Zuckerberg Biohub, San Francisco, CA 94158, USA}
%\address[6]{Instituto de Biocomputaci\'{o}n y F\'{\i}sica de Sistemas Complejos (BIFI), 50009 Zaragoza, Spain}
%\cortext[cor1]{Corresponding author.}

\begin{abstract}
  Understanding thin sheets, ranging from the macro to the nanoscale,
  can allow control of mechanical properties such as
  deformability. Out-of-plane buckling due to in-plane compression can
  be a key feature in designing new materials. While thin-plate theory
  can predict critical buckling thresholds for thin frames and
  nanoribbons at very low temperatures, a unifying framework to
  describe the effects of thermal fluctuations on buckling at more
  elevated temperatures presents subtle difficulties. We develop and
  test a theoretical approach that includes both an in-plane
  compression and an out-of-plane perturbing field to describe the
  mechanics of thermalised ribbons above and below the buckling
  transition.  We show that, once the elastic constants are
  renormalised to take into account the ribbon's width (in units of
  the thermal length scale), we can map the physics onto a mean-field
  treatment of buckling, provided the length is short compared to a
  ribbon persistence length. Our theoretical predictions are checked
  by extensive molecular dynamics simulations of thin thermalised
  ribbons under axial compression.
\end{abstract}
\end{frontmatter}
%\linenumbers

\section{Introduction}
Thin sheets, possibly with embedded kirigami cuts, have been the
object of intense recent study~\cite{Grosso2020}.  A careful design
allows membranes with cuts to stretch far beyond their pristine
limits~\cite{Blees2015, shyu-NatMat-14-785-2015,
  hanakata-Nanoscale-8-458-2016, tang-EML-12-77-2017,
  rafsanjani-PRL-118-084301-2017, hanakata-PRL-121-255304-2018}, to
have non-linear post-buckling
behaviours~\cite{moshe-PRL-122-048001-2019, yang-PRM-11-110601-2018},
and even to exhibit complex motions such as roll, pitch, yaw, and
lift~\cite{dias-sm-48-9087-2017}. Many of these novel effects arise
due to out-of-plane deflections, i.e., escape into the third
dimension.  With such mechanical versatility and straightforward
actuation, kirigami sheets have been used as building blocks for soft
robots, flexible biosensors and artificial
muscles~\cite{rafsanjani-SR-3-7555-2018,
  morikawa-AHM-23-1900939-2019}.  A full theoretical framework for
this rich phenomenology must rest on a thorough understanding of the
fundamental mechanical effects. In particular, out-of-plane motion in
simple kirigami systems (e.g., a sheet with a single slit) have been
described as an Euler buckling
problem~\cite{dias-sm-48-9087-2017}. The buckling of pillars and
plates has been studied for centuries, but a unifying theory to
understand buckling in nanosystems when thermal fluctuations become
important, as in the case of molecularly thin materials such as
MoS$_2$ and graphene~\cite{katsnelson2012graphene}, is still lacking.

In the classical description, the dimensionless F\"oppl-von K\'arm\'an
number vK = $\Ybare W_0L_0/\kappabare$, where $\Ybare$ is the 2D Young's modulus,
$\kappabare$ is the bending rigidity, $W_0$ and $L_0$ are respectively the $T=0$ 
width and length of the ribbon, can be used to
quantify the ease of buckling a thin sheet out of plane at zero
temperature. The picture is more complicated for thermalised
membranes~\cite{Nelson2004}, where $Y$ and $\kappa$ become scale
dependent and, in particular, the bending rigidity is dramatically
enhanced~\cite{Nelson1987,Aronovitz1988,Guitter1989,
  LeDoussal1992,Zhang1993,Bowick1996}. This longstanding theoretical
prediction is consistent with an important study of graphene ribbons
by Blees et al.~\cite{Blees2015}. Using a cantilever
setup, the effective bending rigidity of micron-size graphene at room
temperature was found to increase by a factor of roughly 4000 relative
to the zero-temperature microscopic value. Although it is possible
that some of this increase may be due to quenched random disorder in
the graphene ribbons~\cite{kovsmrlj-PRE-88-012136-2013}, these room
temperature experiments nevertheless demonstrate a striking
enhancement over the $T=0$ density functional theory
predictions~\cite{kudin2001c}. When thermal fluctuations are important,
classical Euler buckling predictions break down. In fact, in
such an entropy-dominated high-temperature setting, some aspects of
nanoribbon behaviours have more in common with linear polymers with
long persistence length~\cite{Kosmrlj2016}.

In this letter, we investigate (i) to what temperature classical Euler
buckling still holds, (ii) how we can locate buckling transitions in
fluctuating ribbons under compression, and (iii) how these buckling
transitions change with temperature and with the ribbon dimensions.
To this end, we develop a mean-field theory (MFT) approach to the
buckling of thermalised ribbon under longitudinal compression and use
molecular dynamics simulations to check our predictions.  The
applicability of our MFT is determined by two crucial length scales:
First, the thermal length
$\lth\sim \kappabare/\sqrt{\Ybare k_\text{B} T}$, where $k_\text{B}$
is the Boltzmann constant, $T$ is the temperature, $\Ybare$ and
$\kappabare$ are the microscopic 2D Young's modulus and bending
rigidity respectively.  And second, the one-dimensional persistence length
$\lp= 2\kappabare W_0/k_\text{B}T$.  We are interested in the regime
$\lth<W_0<L_0<\lp$, where the temperature is high enough that \lth\ is
smaller than the ribbon's width $W_0$, so thermal renormalisation is
significant, but not so high that \lp\ becomes small compared to the
ribbon length $L_0$.

Our theory predicts, and our simulations confirm, that the buckling
transition is delayed, because the renormalised $\YT$ becomes softer and
the renormalised $\kappaT$ becomes stiffer as $T$ increases.  We also
explore the possibility of utilising an out-of-plane uniform
perturbation (e.g., an electric or gravitational field) to break the
height-reversal symmetry. Such fields give an alternative path to
control the buckling transition. Overall, our study provides a new
framework to study buckling in thermalised ribbons which is relevant
to nanomaterials, such as graphene or MoS$_2$, or to biological systems
when the thermal scale is comparable to or less than the system
size. While this work was in progress, we learned of interesting work by
Morshedifard et al.~\cite{morshedifard-arxiv-2020}, who carried out
simulations similar to ours, without, however the introduction of a
symmetry-breaking field, and without the post-buckling mean-field
theory used here.
\begin{figure}[t]
\includegraphics[width=8cm]{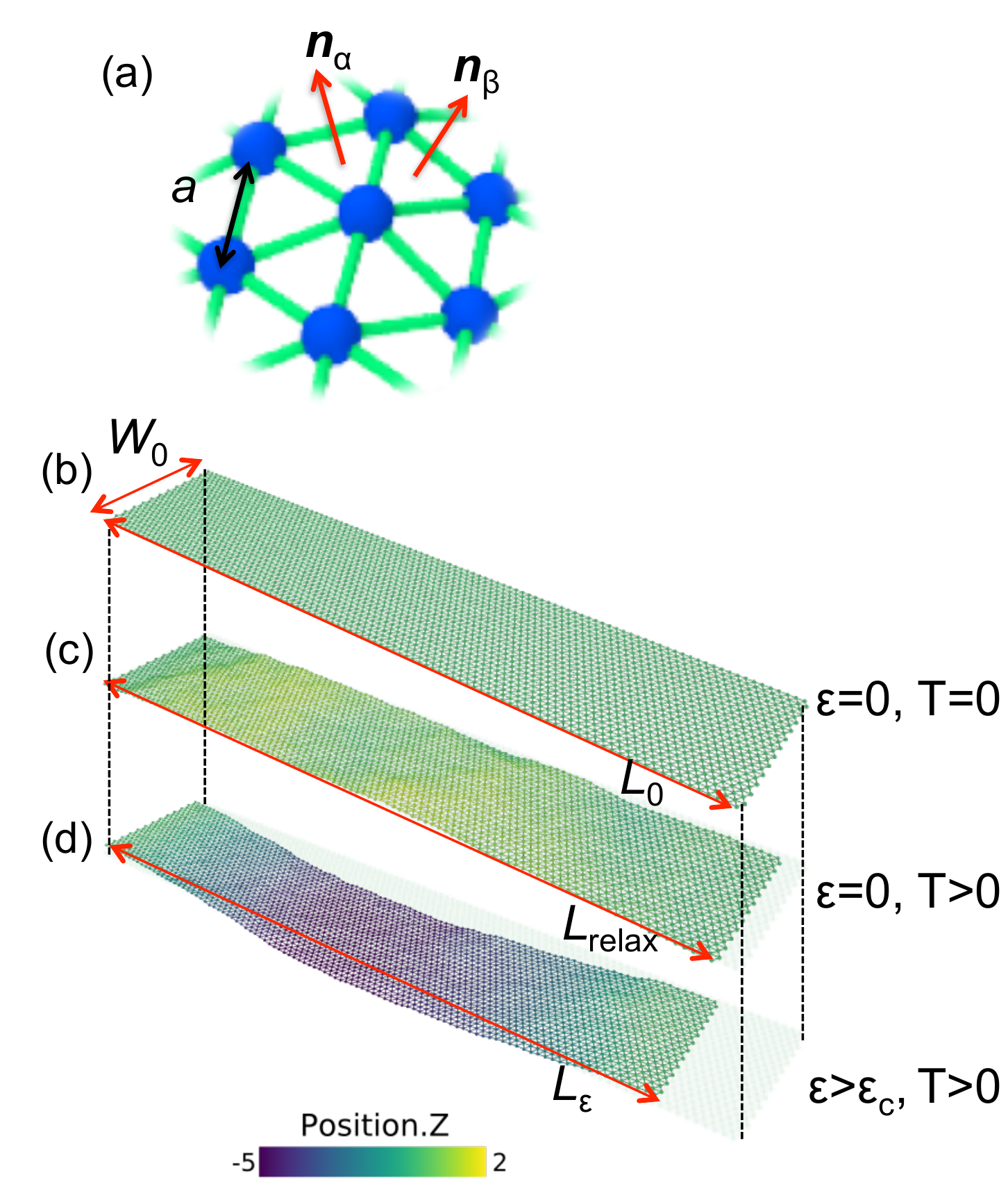}
\caption{(a) Schematic of nodes on a triangular lattice. (b)  Ribbon
 in its flat rest configuration at $T=0$, with length $L_0$ and width $W_0$. 
 (c) Relaxed ribbon at $T>0$ and zero compression. The thermal
  ripples cause the ribbon to shrink from its rest length $L_0$
  down to a thermal projected length $L_\text{relax}$.
  (d) Ribbon at a non-zero temperature and compressed
  beyond the critical buckling strain $\epsilon_\text{c}$. 
  In panels (b)--(d) each end of a ribbon is clamped to have width $W_0$ and the colour map shows the
  height of each node relative to $z=0$ at the two ends. We used OVITO software to visualise the
  ribbons~\cite{ovito}.
  \label{fig:fig1}}
\end{figure}
\section{Model and methods}
\subsection{Coarse-grained model}
We consider a rectangular sheet of size $L_0\times W_0$, with
$L_0 > W_0$, which is discretised by a triangular lattice of
unbreakable bonds, in the crystalline membrane
paradigm~\cite{Bowick2001}.  The triangular lattice used here can be
considered as a convenient dual representation to the honeycomb
lattice usually employed to model graphene. We use the notation $L_0$
to distinguish the $T=0$ rest length from the projected length after
thermal shrinking or compression.  Neighbouring nodes are connected by
harmonic springs and there is an energy cost when the normals ($\pmb{n}_{\alpha}$) of
neighbouring plaquettes are not aligned.  The total energy is given by
\begin{equation} 
 \mathcal H=\frac{k}{2}\sum_{\langle i,j\rangle}(|\pmb{r}_{i}-\pmb{r}_j|-a)^2+\kappasim\sum_{\langle\alpha, \beta\rangle}(1-\pmb{n}_{\alpha}\cdot\pmb{n}_{\beta})  \label{eq:model}
\end{equation}
where $k$ is the harmonic spring constant, $\kappasim$ is the microscopic
bending rigidity and $a$ is the preferred length between two neighbouring
nodes which also sets our unit of length. The first sum is over neighbouring nodes and the second
over neighbouring triangular plaquettes. A schematic is shown in Fig.~\ref{fig:fig1}(a).
Our discretised bare elastic constants are related to the bare continuum
ones by $\kappabare= \sqrt3 \kappasim/2$ and $\Ybare =2k/\sqrt3$~\cite{Seung1988}.

\subsection{Parameters and length scales}
Since we are interested in relatively narrow ribbons, we use
$L_0\sim100a$ and $W_0\sim20a$ (2500 nodes).  Following previous
work~\cite{bowick-PRB-95-104109-2017, Yllanes2017, Yllanes2019}, we
set $k= 1440\kappasim/a^2$, which gives us a F\"oppl-von
K\'arm\'an number of vK $\sim 10^6$, comparable to micron-size 2D
materials such as graphene and MoS$_2$.

As we change the temperature, keeping $k/\kappasim$ fixed, two
crucial length scales, the thermal and persistence
lengths, will vary~\cite{Kosmrlj2016,Yllanes2017}: 
\begin{align}
\lth &= \sqrt{\frac{\pi^3 64\kappabare^2}{3k_\text{B}T\Ybare}}, \label{eq:lth}\\
\lp &= \frac{2\kappabare W_0}{k_\text{B} T}. \label{eq:lp}
\end{align}
We want here to adapt the zero-temperature theory to temperatures high
enough for thermal renormalisation to become significant.  The temperature should not, however,  be so high that \lp\ becomes small compared
to $L_0$ (i.e., we stay far away from the ribbon crumpling regime). In
simulations we fixed $L_0, W_0$ and $k/\kappasim$ while varying
$\kappasim$ and $T$. We simulated over a temperature range
$10^{-7}\leq k_\text{B}T/\kappasim \leq 4$ or equivalently
$10^{-2}\lesssim W_0/\lth \lesssim 10^2$.  In the following we shall
use $W_0/\lth$ as the natural variable for the temperature scaling of
the system, and focus on the regime where $W_0>\lth$.
\subsection{Clamped boundary conditions and molecular dynamics simulations}
We use the HOOMD package~\cite{anderson2020hoomd} to simulate
model~\eqref{eq:model} in the NVT ensemble with a Nos\'e-Hoover
thermostat.  In order to study the buckling dynamics, we clamp the
ribbon by fixing the nodes on the first two rows at both ends. We vary
the distance between the clamped edges to induce the desired
strain. Importantly, we thus operate in a constant-strain ensemble.

Because of thermal fluctuations, the ribbon shrinks from from its $T=0$
rest length $L_0$. We define $L_\text{relax}$ as the projected natural
length at which all stress components are zero and define the
incremental compressive strain as
$\epsilon=1-L_\epsilon/L_{\rm relax}$, where $L_\epsilon$ is the
projected length at a given compressive strain $\epsilon$. At finite
$T$ we have therefore the inequalities
$L_{\epsilon>0} < L_\text{relax} < L_0$, illustrated in
Fig.~\ref{fig:fig1}.

Following \cite{Yllanes2017}, we use a timestep
of $\Delta t = 0.0025\tau$ where $\tau$ is the Lennard-Jones
time  $\tau = \sqrt{m a^2/k_\text{B} T}$
and we use natural units of mass and energy  $m=a=1$.
Our clamped systems are simulated in the NVT ensemble for $10^7$ steps,
saving a snapshot every $10^4$ steps.
For each choice of parameters, we simulate 
either $5$ or (more commonly) $10$ independent runs.
We use a jackknife method (see, e.g.,~\cite{Young2015})
to estimate statistical errors.

\section{Theoretical expectations}
\begin{figure}[t]
\includegraphics[width=8cm]{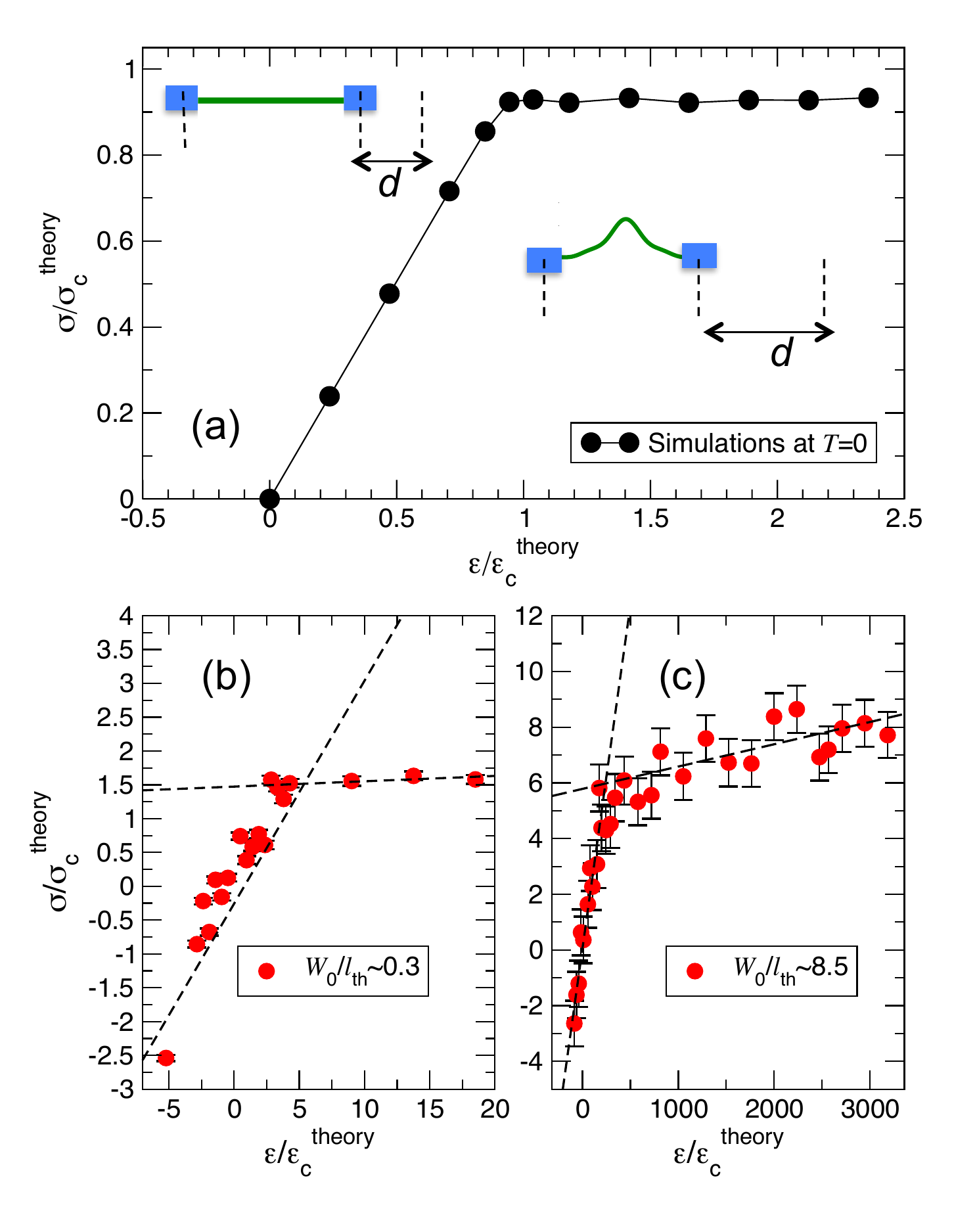}
\caption{(a) Stress as a function of compressive strain scaled by
  their zero-temperature critical values
  ($\sigma_\text{c}^{\rm theory}$ and
  $\epsilon_\text{c}^{\rm theory}$) for a ribbon at (a) $T=0$, (b)
  $W_0/\lth = 0.3$ and (c) $W_0/\lth = 0.3$.  The dotted lines are
  linear fits in the pre-buckling (small-$\epsilon$) regime and in the
  post-buckling regime $\epsilon>\epsilon_\text{c}$. The scaled
  critical stress, which is proportional to the renormalised bending
  rigidity $\kappaT$, increases with increasing $W_0/\lth$ (or
  temperature). In contrast, the slope ($\YT$) becomes smaller with
  increasing $W_0/\lth$. Note that the very different horizontal
  scales in (b) and (c).}
  \label{fig:fig2}
\end{figure}
The most dramatic signature of the buckling transition occurs in
stress-strain curves. Fig.~\ref{fig:fig2} shows the stress as a
function of the strain as measured from our simulations when $T=0$,
then at a low $T$ such that $W_0/\lth=0.3$, and finally at a more
elevated temperature such that $W_0/\lth=8.5$, where thermal
fluctuations have a stronger effect. The computed Young's modulus,
critical stress, and critical strain for $T=0$ are within 10\% of the
theoretical predictions
$[\frac{\Ybare^{\rm simulation}}{\Ybare^{\rm theory}}=0.99,
\frac{\sigma_\text{c}^{\text{simulation}}}{\sigma_\text{c}^{\rm
    theory}}=0.92, \frac{\epsilon_\text{c}^{\text{\rm
      simulation}}}{\epsilon_\text{c}^\text{theory}}=0.94]$.
We attribute the small deviations to our discretised clamped boundary
conditions. The stress-strain curves can be understood via the
following simple argument: we write the energy of a possibly bent
compressed ribbon of width $W_0$, with displacements uniform along the
$y$ direction, as
\begin{equation}
\begin{split}
E/W_0=&\frac{1}{2}\Ybare\int^{L/2}_{-L/2}\left(\frac{\dd u(x)}{\dd x}\right)^2\dd x-\sigma_{xx}d\\
+& \frac{1}{2}\kappabare\int^{L/2}_{-L/2}\left(\frac{\dd^2h(x)}{\dd x^2}\right)^2,\\
\end{split}
\label{eq:G1}
\end{equation}
where $u(x)$ is the displacement field along the $x$ axis, $h(x)$ the
displacement perpendicular to the ribbon, and $-\sigma_{xx}d$
represents the work done by a force $F=W_0\sigma_{xx}$ to compress the
ribbon an amount $d$ along $\hat{x}$ relative to its natural length
$L$. Here, $\Ybare$ and $\kappabare$ are the 2D Young's modulus and bending
rigidity which measure the compressional and bending energies
respectively. In the compressed, but unbuckled, state the strain is
$\epsilon=\frac{\dd u(x)}{\dd x}=d/L$ and from Eq.~\eqref{eq:G1}, the
compressional energy is $E_{\rm comp}=\frac{1}{2}W_0\Ybare d^2/L$.
In this regime, we minimise over $d$ to find Hooke's Law
$\sigma=F/W_0=\Ybare\epsilon$, which accounts for the first, linear part of
the stress-strain curve. Beyond the critical strain
$\epsilon_\text{c}$, however, the system prefers to trade compressional energy for
bending energy. As we shall discuss below, for tangential boundary
conditions at two ribbon ends, as is the case for our simulations, we
have the usual buckling instability when $\epsilon>\epsilon_\text{c}$,
$\epsilon_\text{c}=4\pi^2\kappabare/(\Ybare L^2)$~\cite{Landau}.

What is the incremental stress $\delta \sigma$ associated with an
additional strain $\delta \epsilon=\delta d/L$ when
$\epsilon>\epsilon_\text{c}$? To this end, we assume the compressional
energy vanishes. We can now regard $x$ as a coordinate embedded in the
ribbon. Note that the tipping angle $\theta(x)$ of the normal away
from the $z$-axis is given by $\theta(x)\approx\frac{\dd h}{\dd x}$,
so that the additional energy associated with the buckled state can be
rewritten as
\begin{equation}
\Delta E/W_0\approx \frac{1}{2}\kappabare\int_{-L/2}^{L/2}\left(\frac{\dd\theta(x)}{\dd x}\right)^2\dd x-\delta\sigma d,
\label{eq:postbuckling}
\end{equation}
where $-\delta\sigma W_0d$ is the extra work done beyond the buckling
transition by the stress increment $\delta\sigma$. Once buckling leads
to a ribbon with a well-developed looping arch, i.e., $\epsilon\gg \epsilon_\text{c}$
we expect that $\frac{\dd\theta(x)}{\dd x}\sim\frac{\pi d}{L^2}$ so that
the normal turns  an angle $\delta \theta\sim\pi/2$ when
$d\sim L/2$. The energy associated with Eq.~\eqref{eq:postbuckling} is
then $\Delta E\sim W_0\kappabare d^2/L^3$. Upon minimising this
expression over $d$, we obtain
\begin{equation}
\delta\sigma_{xx}=\text{const.}\frac{\kappabare d}{L^3}\approx\text{const.}\frac{\kappabare}{L^2}\delta\epsilon.
\end{equation}
Thus, the slope of the stress-strain curve beyond $\epsilon_\text{c}$, once
the buckling transition becomes well developed, should be of the order
$\kappabare/ L^2$, as might have been guessed from dimensional
analysis.

We conclude that the ratio of the pre- and post-buckling slopes is
$\sim \Ybare L^2/\kappabare$, i.e., it is of the order the F\"oppl-von K\'arm\'an
number $\sim10^6$ in our simulations! Hence, it is not surprising that
the zero-temperature stress-strain curve looks nearly flat in
Fig.~\ref{fig:fig2}(a).\footnote{Both the critical strain
  $\epsilon_\text{c}\sim\kappabare/(\Ybare L^2)$ and the post-buckling slope
  $\sim\kappabare/L_0^2$ vanish in the thermodynamic limit
  $L_0\rightarrow\infty$.} There is, however, a hint of a non-zero
slope at finite temperatures in our simulations when $W_0/\lth=0.3$,
which becomes more pronounced when $W_0/\lth$=8.5. As discussed below,
we attribute this enhanced post-buckling slope to a strong
$W_0$-dependent upward renormalization bending rigidity
$\kappabare\rightarrow \kappaT$, due to thermal fluctuations. Moreover, by
rescaling the stress and the strain with their respective
zero-temperature critical buckling compression and strain, we can see
that the critical strain and critical buckling compression increase
with increasing $T$, or equivalently increasing $W_0/\lth$, as shown
in Fig.~\ref{fig:fig2}(b) and (c).

The argument above cannot tell us the details of what happens close to
$\epsilon_\text{c}$, where one must account for delicate balance between
compression and bending energies. To understand this regime, we  now
construct a simple Landau-like theory of the buckling transition,
appropriate to the constant-strain ensemble enforced by our constant
NVT simulations.

\subsection{Mean-field theory}
As the ribbon is compressed along the longitudinal $x$ direction it
can both compress and deflect out of plane in the $z$ direction. We
work in the Monge representation and denote the vertical displacement
by $h(x,y)$. In this derivation we denote the \emph{instantaneous}
projected length after a compression $d$ (to produce a dimensionless
compressive strain $\epsilon$) by $L_\epsilon$.  To control the
buckling order parameter, we also impose an out-of-plane electric
field $\mathcal E$ coupled to the height of a charged ribbon,
generating a potential energy
$V_{\perp}=-\int_{-L_\epsilon/2}^{L_\epsilon/2}\rho\mathcal{E}\
h(\vec{x})\ \dd^2 x$,
where $\rho=Q/(L_0W_0)$ is the charge density. To describe a ribbon in a
gravitational field we simply need to substitute $\rho=m/(L_0W_0)$ and
$\mathcal{E}=g$.  We assume a large F\"oppl-von K\'arm\'an number
$\Ybare L_0W_0/\kappabare$ (easily achieved for graphene and Mo$S_2$),
in which the stretching along the ribbon will be comparatively
small. The total free-energy cost is given by
\begin{equation}
\begin{split}
G&=\frac{1}{2}\int \dd^2 x\left[\left(\nabla^2 h\right)^2+2\mu u^2_{ij}+\lambda u^2_{kk} \right]-\rho\mathcal{E}\int \dd^2 x h\\
&-\sigma\int \dd^2x (\partial_x u_x) , \\
\end{split}
\label{eq:G2D}
\end{equation}
where
$u_{ij}=(\partial_i u_j+\partial_j u_i)/2+(\partial_i
h)(\partial_jh)/2$
and $\sigma_{xx}= \sigma$ denotes a uniaxial stress at the clamped
edges. Notice that, since the centre-of-mass height is
$h_{\text CM}=\frac{1}{W_0L_0}\int \dd^2x h(\vec{x})$, we can write
$G=G_0-\mathcal{E}Qh_{\text{CM}}$ and the thermally averaged
centre-of-mass height $h_{\text{CM}}$ in the full fluctuating which we
are only approximating here
\begin{equation}
\langle h_{\text{CM}}\rangle=\frac{1}{Z}\int\mathcal{D}[h, u_i]\ h_{\rm CM}\ee^{-(G_0-\mathcal{E}Qh_{\text CM})/k_{\text B}T},
\label{eq:exp_hcm}
\end{equation}
where $Z=\int\mathcal{D}[h, u_i]\ \ee^{-E/k_{B}T}$ is the partition
function. Since we are interested in the buckling response due to an
external field, we also study the height susceptibility defined as
$\chi=\dd\langle h_{\text CM}\rangle/\dd\mathcal{E}$. Upon using Eq.~\eqref{eq:exp_hcm} we obtain 
\begin{equation}
\chi \propto \langle h^2_\text{CM}\rangle - \langle h_\text{CM}\rangle^2.
\label{eq:Xi_hcm}
\end{equation}

We can further simply the physics into a 1D buckling problem. We
approximate $h(x,y)\approx h(x)$ and define charge density
$\rho=Q/L_0$, an effective 1D bending rigidity and Young's modulus
given by $\kappa_{\rm 1D}=\kappabare W_0$ and $Y_{\rm 1D}=\Ybare W_0$,
respectively, where $\kappabare$ and $\Ybare$ denote $T=0$ values of
the elastic constants. Within a Monge representation, we can
approximate
$L_{\epsilon}+d \simeq
L_{\epsilon}+\frac{1}{2}\int_{-L_\epsilon/2}^{L_\epsilon/2}
\left(\frac{\dd h}{\dd x}\right)^2\ \dd x$,
where the strain $\epsilon$ is given by $\epsilon=d/L_{\rm relax}$.
The total energy then consists of bending, stretching and work done by
the external compressive force $F$ and an out-of-plane field,
\begin{equation}
\begin{split}
 G[h, \mathcal{E}]&=\frac{\kappa_{\rm 1D}}{2}\int_{-L_\epsilon/2}^{L_\epsilon/2}\dd x\left(\frac{\dd^2h}{\dd x^2}\right)^2\\
&\quad +\frac{Y_{\rm 1D}}{2L_{\epsilon}}\left[\int_{-L_\epsilon/2}^{L_\epsilon/2}\dd x\frac{1}{2}\left(\frac{\dd h}{\dd x}\right)^2\right]^2\\
&\quad -\frac{F}{2}\int_{-L_\epsilon/2}^{L_\epsilon/2}\dd x\left(\frac{\dd h}{\dd x}\right)^2-\rho\mathcal{E}\int_{-L_\epsilon/2}^{L_\epsilon/2} h\ \dd x.
\end{split}
\label{eq:G}
\end{equation}
Note that we have eliminated, or ``integrated out'', the in-plane
phonons. See~\ref{sec:variational} for a detailed derivation
of Eq.~\eqref{eq:G}, which incorporates our constant-strain boundary
conditions. Note also the non-local character of the second,
stretching term.  Lifshitz and Cross~\cite{lifshitz2008nonlinear} have
described equations of motion for micro-electromechanical devices with
a similar non-local term. The ansatz of the first buckling mode
$h(x)=\frac{1}{2}h_{\rm M}\left[1+\cos \left(\frac{2\pi
      x}{L_{\epsilon}}\right)\right]$,
which allows for a height $h_\text{M}$ midway between the clamps and
satisfies the boundary conditions
$\left. \frac{\dd h}{\dd x}\right|_{x=\pm L_{\epsilon}/2}=0$, then
leads to an expansion in the buckling amplitude $h_{\text M}$
\begin{equation}
  G=\frac{\pi^2}{4L_{\epsilon}}\left(\frac{4\kappa_{\rm 1D}\pi^2}{L_{\epsilon}^2}-F\right)h^2_{\rm M}+\frac{\pi^4Y_{\rm 1D}}{32L_{\epsilon}^3}h^4_{\rm M}-\frac{\rho L_{\epsilon} \mathcal{E}}{2}h_{\rm M}. \label{eq:Gansatz}
\end{equation}
Note that, although Eq.~\eqref{eq:Gansatz} resembles a Landau theory
near a critical point, the expansion parameter depends in a
non-trivial way on the system dimension $L_\epsilon$. Note also that
the single mode approximation only makes sense close to the
transition; many more Fourier modes would be required to describe the
fully developed post-buckling looping arch that develops for large
strains, as in Fig.~\ref{fig:fig2}(a).
\subsection{Euler buckling at $T=0$}
For $\mathcal{E}=0$, we can minimise Eq.~\eqref{eq:Gansatz} over $h_{\text M}$ to obtain a
critical 2D compressive stress of
$\sigma_\text{c}=4\pi^2\kappabare/L_{\epsilon_\text{c}}^2$, and a corresponding
critical buckling strain
$\epsilon_{\rm
  c}=\sigma_\text{c}/\Ybare=\frac{4\pi^2\kappabare}{\Ybare L_{\epsilon_\text{c}}^2}$,
where $L_{\epsilon_\text{c}}$ is the projected length at the critical
buckling strain.\footnote{For a very large F\"oppl-von K\'arm\'an
  number vK we can approximate $L_{\epsilon}$ as $L_0$} These are
the critical load and critical strain of classical Euler buckling with
tangential boundary conditions~\cite{Landau}. The buckling amplitude
is then
\begin{equation}
h_\text{M} = \begin{cases}
0, & \epsilon < \epsilon_\text{c} \text{ (or $\sigma < \sigma_\text{c}$)},\\
\pm\frac{2L_{\epsilon_\text{c}}}{\pi}\sqrt{\epsilon-\frac{4\kappabare\pi^2}{\Ybare L_{\epsilon_\text{c}}^2}}, &
\epsilon \geq  \epsilon_\text{c} \text{ (or $\sigma \geq \sigma_\text{c}$)}.
\end{cases}
\end{equation}
To test the above approach, we compared simulations at $T=0$ with the analytical predictions.
 These simulations reproduced the
square-root scaling predicted by the theory and yielded consistent
values for the Young's modulus, critical stress and critical strain
(see~\ref{sec:T0} for details and plots).
\subsection{Response function near critical buckling}
At the critical point the system becomes sensitive to external perturbation. In
analogy with the magnetic susceptibility of  an Ising system, within the MFT we
can define a height susceptibility as the linear response to a uniform
out-of-plane external field,
\begin{align}\label{eq:chi}
\chi\equiv\left. \frac{\partial h_{\rm M}}{\partial \mathcal{E}}\right|_{\mathcal{E}=0} &=
  \begin{cases}
    \frac{Q}{\Ybare\pi^2}\frac{L_{\epsilon_\text{c}}}{W_0}(\epsilon_{\rm c}-\epsilon)^{-1}    & \quad \text{if } \epsilon<\epsilon_{\rm c}\\
    \frac{Q}{2\Ybare\pi^2}\frac{L_{\epsilon_\text{c}}}{W_0}(\epsilon-\epsilon_{\rm c})^{-1} & \quad \text{if } \epsilon>\epsilon_{\rm c}.
  \end{cases}
\end{align}
This response function diverges at the buckling transition. Hence, the system
becomes infinitely sensitive to the out-of-plane field $\mathcal{E}$
as the buckling transition is approached. Note also that $\chi$ is
larger for small aspect ratio $W_0/L_{\epsilon}$. Eq.~\eqref{eq:Gansatz}
predicts a non-linear dependence of the buckling amplitude $h_{\text M}$ on  $\mathcal{E}$
when $\epsilon=\epsilon_\text{c}$
\begin{equation}
h_{\text M}=L_{\epsilon_\text{c}}\left(\frac{4Q\mathcal{E}}{\pi^4\Ybare W_0}\right)^{1/3}\,.
\label{eq:hm_efield}
\end{equation}
The finite-temperature generalisation of this susceptibility is given by Eq.~\eqref{eq:Xi_hcm}.
\subsection{Thermalised Euler buckling}
As the temperature increases and the thermal length \lth (see Eq.~\eqref{eq:lth})
becomes smaller than the membrane's dimensions, the elastic 
constants of the system are renormalised. For ribbons with $W_0<L_0$
this renormalisation is cut off by the width 
and leads to the following renormalized elastic constants~\cite{Kosmrlj2016}:
\begin{align}
\kappaT(W) &\simeq
  \begin{cases}
    \kappabare     & \quad \text{if } W_0< \lth,\\
    \kappabare\left(\frac{W_0}{\lth}\right)^{\eta}  & \quad \text{if } W_0> \lth,
  \end{cases}\label{eq:WR}\\
\YT(W) &\simeq
  \begin{cases}
    \Ybare    & \quad \text{if } W_0<\lth\\
    \Ybare\left(\frac{W_0}{\lth}\right)^{-\eta_u}  & \quad \text{if } W_0>\lth,
  \end{cases}\label{eq:YR}
\end{align}
where $\eta\approx0.8-0.85$ and $\eta_u\approx0.3-0.4$ from analytical
computations~\cite{LeDoussal1992, kownacki-PRE-79-040101-2009,
  Kosmrlj2016} and molecular dynamics or Monte Carlo
simulations~\cite{Bowick1996,los-PRB-80-121405-2009, roldan-PRB-83-174104-2011,
  bowick-PRB-95-104109-2017}. We expect, therefore, a strongly $W_0$
and temperature-dependent stiffening in the bending rigidity and
softening in the Young's modulus. Upon substituting the renormalised
elastic constants in to the MFT, we obtain a scaling for the critical
2D stress of $\sigma_\text{c}\propto (W_0/\lth)^{\eta}$ and a critical
buckling strain $\epsilon_\text{c}\propto (W_0/\lth)^{\eta+\eta_u}$.  Because
$\lth \propto T^{-1/2}$ and using $\eta\approx0.8$
and the scaling relation~\cite{Aronovitz1988} $2\eta+\eta_u=2$, we 
see that  $\sigma_\text{c}$ and $\epsilon_\text{c}$ are predicted to increase
with increasing temperature with non-trivial power laws,
$\epsilon_\text{c}\sim T^{0.6}$ and $\sigma_\text{c}\sim T^{0.4}$.

\section{Numerical results for finite temperature}
The MFT section explains how we can use the maximum height
$h_\text{M}$ of a compressed ribbon as an order parameter for a
buckling transition and estimate how the critical strain and critical stress will
shift with increasing temperature. We now test this theoretical
prediction against molecular dynamics simulations.  We use the
notation $\langle O\rangle$ for the average in the $NVT$ ensemble of
the observable $O$.

It will be convenient to replace $h_\text{M}$ by the height of the
ribbon centre of mass $h_{\rm CM}=\frac{1}{N}\sum_ih_i$ as our order
parameter. There is, however, a subtle point to be considered. In the
absence of an external field ($\mathcal{E} = 0$) our
energy~\eqref{eq:G} is invariant with respect to height-inversion
symmetry. This means that configurations with $h_\text{CM} = \pm h$
have the same probability and would seem incompatible with the result
$\langle h_\text{CM}\rangle \neq 0$ for
$\epsilon > \epsilon_\text{c}$.  As with conventional magnetic phase
transitions, this apparent paradox is resolved by realising that, in the
limit of large system sizes, the system undergoes spontaneous symmetry
breaking~\cite{zinn-justin:05}. Formally, we could consider a small
symmetry-breaking field $\mathcal{E}$ to establish a preferred
direction and take the double limit
\begin{equation}
\langle h_\text{CM}\rangle = \lim_{\mathcal E \to0}\lim_{A\to\infty} \langle h_\text{CM}\rangle_A,
\end{equation}
where $A$ denotes the system size.  Notice that if we reversed the
order of the limits $\langle h_\text{CM}\rangle$ would always
vanish. This is the situation in any computer simulation, where flips
between the up and down states are always possible after a finite long
time. The metastable dynamics for $\mathcal{E}=0$ and the behaviour of
the flipping time for a molecular dynamics simulation will be
considered in a future work~\cite{sourav}.

The previous discussion is in complete analogy to the magnetisation
$m$ of a magnetic system, where $m$ plays the role of our height
variable. In~\ref{sec:stressstrain} we explore the 
 behaviour of the
susceptibility
$\chi=\left. \frac{\dd h_\text{CM}}{\dd\mathcal{E}}\right|_{\mathcal{E}=0}$ via
  simulations and find that these fluctuations become very large as
  the buckling transition is approached from below.

\subsection{Buckling induced by an external field}
\begin{figure}[t]
\includegraphics[width=9cm]{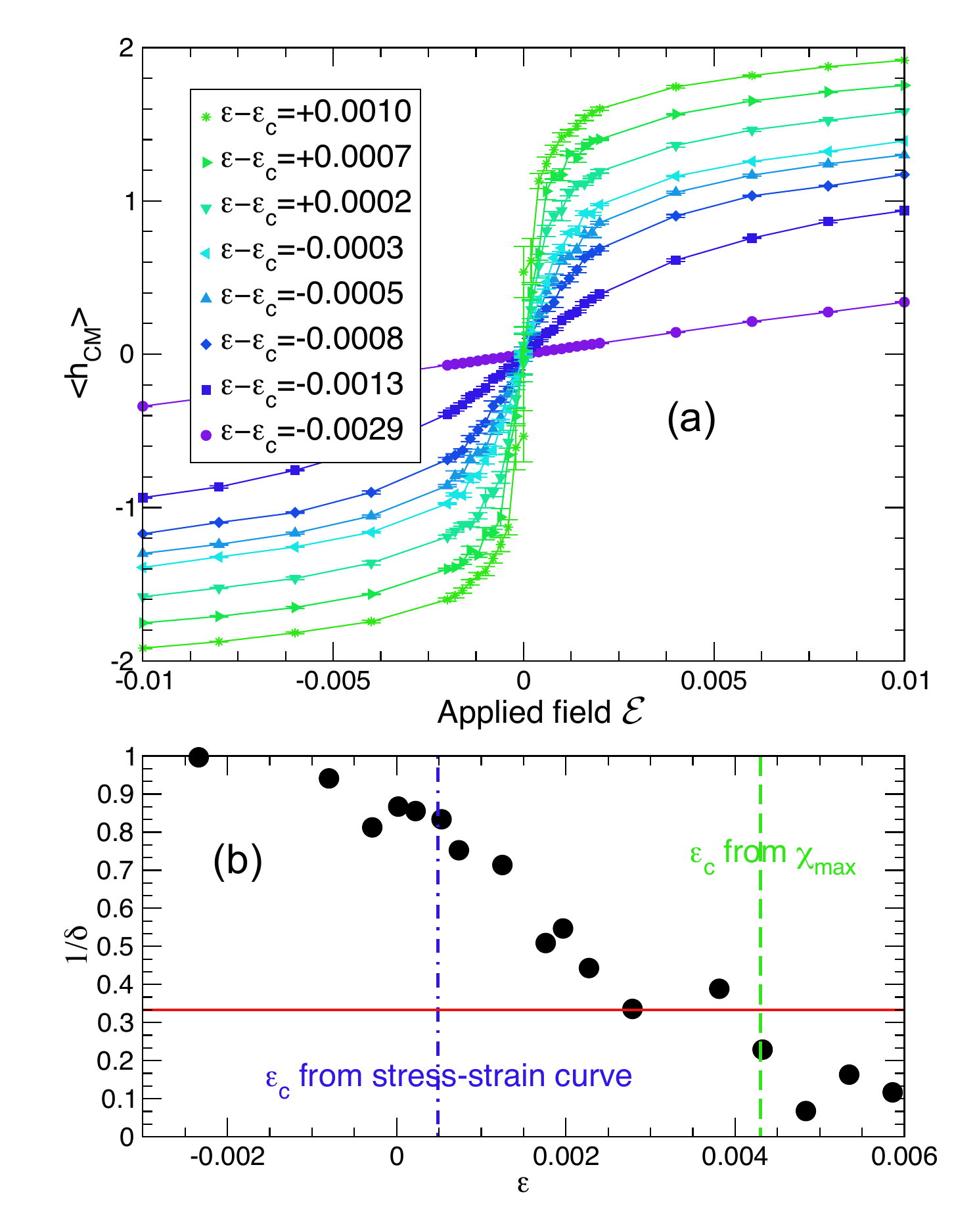}
\caption{(a) The height centre of mass $\langle h_{\rm CM}\rangle$ as
  a function of the out-of-plane field $\mathcal{E}$ for different
  strains relative to critical strain $\epsilon_\text{c}$ obtained from the
  stress-strain curve. The slope (susceptibility) increases closer to
  the buckling transition. (b) The exponent $1/\delta$ as a function
  of $\epsilon$. The critical strain obtained from the height
  susceptibility $\chi$ (green vertical dashed line) does not coincide
  with the $\epsilon_\text{c}$ obtained from the stress-strain curve (blue
  vertical dashed line). We see that $h_\text{CM}\propto \mathcal{E}$ far
  below the buckling transition and that it becomes more sensitive (smaller
  $1/\delta$) as the system becomes closer to the transition. The
  $1/\delta$ exponent is close to 1/3 (red horizontal line) when
  $\epsilon$ is close to the value when $\chi$ is at maximum.}
\label{fig:fig3}
\end{figure}
The definition of the broken-symmetry phase becomes difficult for
finite sizes, since the ribbon can always flip between the up and down
states.  We can break this degeneracy by applying an external field
perpendicular to the plane. From Eq.~\eqref{eq:hm_efield} we expect
steep curves of $\langle h_\text{CM}\rangle_\mathcal{E}$ as
$\mathcal{E}\rightarrow 0$, near the buckling transition, or
equivalently
$\frac{\dd h_{\rm CM}}{\dd\mathcal{E}}\propto\mathcal{E}^{-2/3}$.

We can test this prediction in MD simulations by changing the
perturbing field for compressions at a constant
temperature. Specifically, we simulate ribbons with $W_0/\lth\sim8.5$
($k_\text{B}T/\kappasim=0.05$) and apply an $\mathcal{E}$ up to
0.01. To save computing time we only simulated $\mathcal{E}>0$. In
Fig.~\ref{fig:fig3}(a), we see that well below the buckling transition
the average centre-of-mass height $\langle h_{\rm CM}\rangle$ is
weakly dependent on the field. As we approach the critical buckling
strain,
$\left. \frac{\dd h_\text{CM}}{\dd\mathcal{E}}\right|_{\mathcal{E}=0}$
becomes larger. Along the iso-strain $\epsilon=\epsilon_\text{c}$
where $\chi$ is at maximum, we expect
$\langle h_\text{CM}\rangle\propto \mathcal{E}^{1/\delta}$ where
$\delta=3$ (see Eq.~\ref{eq:hm_efield}). We can fit our data to
calculate exponent $\delta$.  In Fig.~\ref{fig:fig3}(b) we also plot
the critical strain obtained from stress-strain curve and from the
peak of the height susceptibility $\chi$. Interestingly, we find a
proportionality between the critical strains obtained from
stress-strain curves and critical strains obtained from the peaks of
$\chi$; however, these two values do not coincide exactly (see~\ref{sec:variational} for
more details). We find that $1/\delta$ is close to 1/3 as $\epsilon$
approaches $\epsilon_\text{c}$, where $\chi$ is maximum. We hope to
investigate this proportionality in future work. Similar to
magnetic-based memories, one could use the up and down buckling in a
double-clamped ribbon to store information, which can be controlled by
compression $\epsilon$, temperature $T$, or perturbing out-of-plane
field $\mathcal{E}$.
\subsection{The centre-of-mass height behaviour under compression}
\begin{figure}[t]
\includegraphics[width=9cm]{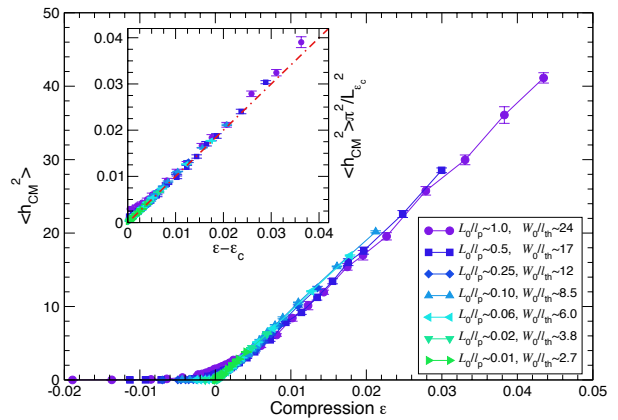}
\caption{Average of the squared centre-of-mass height
  $\langle h_{\rm CM}^2\rangle$ as a function of compressive strain
  $\epsilon$ when $\mathcal{E}=0$. In
  the pre-buckling region $\langle h_{\rm CM}^2\rangle$ is close to zero, whereas in
  the post-buckling region $\langle h_{\rm CM}^2\rangle$ goes linearly with
  $\epsilon$. The inset shows the dimensionless order parameter
  ${\langle h_{\rm CM}^2\rangle \pi^2}/{L_{\epsilon_\text{c}}^2}$ as a function of
  $\epsilon-\epsilon_\text{c}$. The collapse of all data with a slope of one,
  as in Eq.~\eqref{eq:hcm2T}, agrees with our mean-field theory.}
\label{fig:fig4}
\end{figure}
As we discussed earlier we can locate the buckling transitions from
stress-strain curves using data like those in Fig.~\ref{fig:fig2}. We
expect these curves will have a constant slope close to $\epsilon=0$,
given by the Young's modulus, and another slope
$\sim\kappaT/L_\epsilon^2$ beyond the buckling point. The crossing
point of the pre- and post-buckling curves gives the critical buckling
load $\sigma_\text{c}$ and critical strain $\epsilon_\text{c}$ (see~\ref{sec:stressstrain}
for more details).

To provide a more quantitative test of the MFT, we can
use the relation
$h^2_{\rm
  CM}=\left(\frac{1}{L_\epsilon}\int^{L_\epsilon/2}_{-L_\epsilon/2}h\
  \dd x\right)^2=\frac14{h^2_{\rm M}}$
to define a dimensionless buckling parameter at $T=0$:
\begin{equation}
\frac{h_{\rm CM}^2 \pi^2}{L^2_{\epsilon_\text{c}}{(T=0)}}=\epsilon-\epsilon_\text{c}{(T=0)},
\label{eq:hcm2}
\end{equation}
where $\epsilon_\text{c}{(T=0)} =\frac{4\kappabare\pi^2}{\Ybare L^2_{\epsilon_\text{c}}{(T=0)}}$.
At finite temperature, we expect the same relation to hold,
with the corresponding $\epsilon_\text{c}$ given by the renormalised
constants:
\begin{equation}
\frac{\langle h_{\rm CM}^2\rangle \pi^2}{L^2_{\epsilon_\text{c}}}=\epsilon-\epsilon_\text{c}.
\label{eq:hcm2T}
\end{equation}
Note that, at finite $T$, $L_{\epsilon_\text{c}}$ and $L_{\rm relax}$ are
temperature dependent.  Fig.~\ref{fig:fig4} shows
$\langle h_{\rm CM}^2\rangle$ as a function of $\epsilon$ for
different $W_0/\lth$. The linear dependence is clear.  To test the MFT
prediction, we subtract $\epsilon_\text{c}$, found previously from the
stress-strain curve analyses, from $\epsilon$. Remarkably, we indeed
find a data collapse with a slope of one for
$\epsilon>\epsilon_\text{c}$, in accordance with MFT and Eq.~\eqref{eq:hcm2T}.
At high temperatures, however, the transitions grow less sharp,
presumably due to finite-size effects.

\subsection{The renormalised elastic constants}
\begin{figure}[t]
\includegraphics[width=9cm]{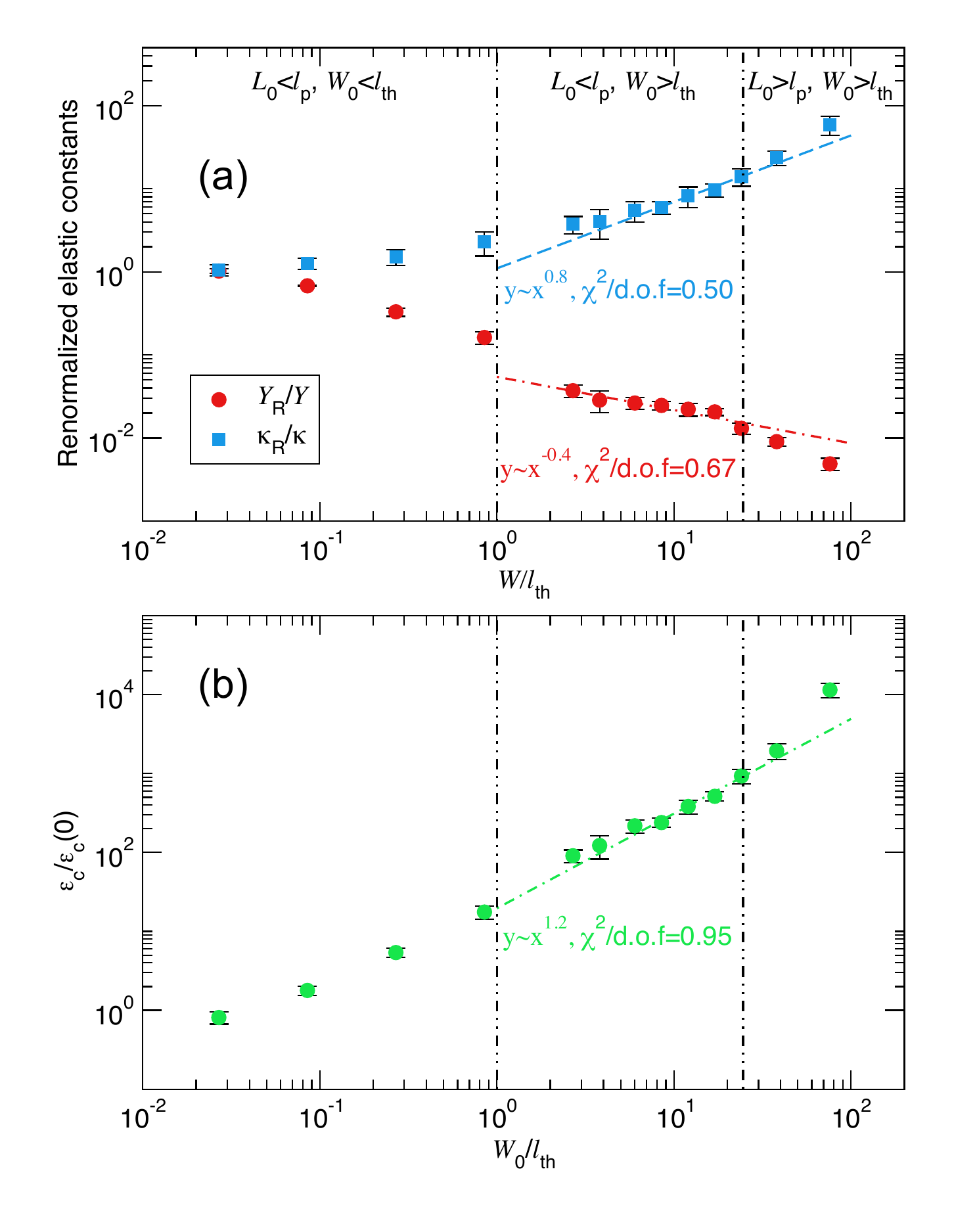}
\caption{(a) Young's modulus $\YT$, bending rigidity $\kappaT$ and (b)
  critical strain $\epsilon_\text{c}$ as a function of $W_0/\lth$. 
  The expected theoretical scaling of Eq.~\eqref{eq:fit} is 
  an excellent fit in the regime  $W_0>\lth, L_0<\lp$
  with $\eta=0.8$ and $\eta_u=0.4$.
}
\label{fig:fig5}
\end{figure}
Next we examine how the elastic constants and critical buckling change
with temperature.  We plot $\kappaT$, $\YT$, and $\epsilon_\text{c}$,
obtained from MD simulations, as a function of $W_0/\lth$ in
Fig.~\ref{fig:fig5}.  At very low temperatures, when
$W_0/\lth\ll 1$ and $L_0/\lp\ll 1$, these three parameters 
approach their zero-temperature values.
In this regime
thermal fluctuations are weak, and thus our system behaves like a
classical ribbon. In the $W_0/\lth>1$ regime, on the other hand, we see stiffening in
$\kappaT$ and softening in $\YT$.
We test Eqs.~\eqref{eq:WR} and~\eqref{eq:YR}
by fitting our data
for  $W_0/\lth>1$ and $L_0/\lp<1$  to 
the following expressions:
\begin{align}
\frac{\YT}{\Ybare}&= A_Y x^{-\eta_u}, &
\frac{\kappaT}{\kappabare}&= A_\kappa x^{\eta}, &
\frac{\epsilon_\text{c}}{\epsilon_\text{c}^{T=0}}&= A_\epsilon x^{\eta+\eta_u},
\label{eq:fit}
\end{align}
where $x= W_0/\lth$.

We first set  the exponents to their
expected values $\eta=0.8$ and $\eta_u=0.4$ and fit only the $A_i$ to check
for consistency. The fits are excellent for the three quantities, with $\chi^2$
goodness-of-fit estimators per degree of freedom of
$\chi^2_Y/\text{d.o.f.} = 4.04/6$,
$\chi^2_\kappa/\text{d.o.f.} = 3.01/6$ and
$\chi^2_\epsilon/\text{d.o.f.} = 5.69/6$.

We have also tried to compute the exponents independently with fits to
Eq.~\eqref{eq:fit} without restricting their values. This is a
difficult computation~\cite{bowick-PRB-95-104109-2017}, since the
range of $W_0/\lth$ that can be accessed in thermalised simulations is
limited.  We have, however, obtained reasonable estimates of
$\eta_u=0.41(10)$ and $\eta=0.67(18)$. The Young's modulus $\YT$ softens
as the ribbon length $L_0$ becomes comparable to the persistence
length \lp. Very recent work by Morshedifard et al. also found an
increase in buckling load of square sheets with increasing
temperature~\cite{morshedifard-arxiv-2020}. It has also been shown in Ref.~\cite{jiang2014buckling}
that the critical buckling strain of Mo$S_2$ sheets (described by a
Stillinger-Weber potential) increases with increasing
temperature. To summarise, in the
semi-flexible regime where $L_0<\lp$ and $\lth<W$ we find that the
mechanics of thin ribbons becomes temperature dependent with
$\YT\propto T^{-\eta_u/2}$, $\kappaT\propto T^{\eta/2}$, and
$\epsilon_\text{c}\propto T^{(\eta_u+\eta)/2}$.

\section{Conclusions}
In this letter we demonstrate that the buckling of thermalized
ribbons, when studied via molecular dynamics simulations, can be described
by a mean-field theory with renormalized elastic constants when the
ribbon length is shorter than the persistence length. We provide three
independent ways of locating the buckling transition.  In the first
approach we use the stress-strain curve to locate buckling and indeed
find that the buckling is \emph{delayed} with increasing
temperature. The second approach is via height fluctuations (\ref{sec:stressstrain}),
in analogy with the study of susceptibility in magnetic systems.
Such an increase in height fluctuations close to the buckling
transition was recently observed in the study of buckling of 1D
colloidal systems~\cite{stuij-PRR-2-023033-2019}.  Lastly, we find that
the height becomes highly sensitive to an out-of-plane symmetry-breaking
field $\mathcal{E}$ close to the transition.

While the buckling transitions of thermalised nanoribbons and phase
transitions in magnetic systems seem to share similar behaviours, the
critical buckling strain is system-size dependent
($\epsilon_{\text c}\propto1/L^2$), whereas the critical temperature
of a magnetic system is typically independent of system size. Our
simulations suggest regions in which the mean-field theory
approximately holds. These regions are determined by the ratio between
the system sizes ($L_0, W_0$) and the relevant thermal lengths
($\lth, \lp$). In the low temperature regime ($L_0<\lp$ and
$\lth>W_0$), the classical (zero temperature) plate theory holds. In
the intermediate (semi-flexible) regime where $L_0<\lp$ and $\lth<W_0$
we find that the mechanics of thin ribbons with fixed width $W_0$ can
be described with a mean-field theory with temperature dependent
elastic constants $\YT\propto T^{-\eta_u/2}$,
$\kappaT\propto T^{\eta/2}$.

Because of the softening in $\YT$ and stiffening in $\kappaT$, the
buckling threshold increases with temperature,
$\epsilon_\text{c}\propto T^{(\eta_u+\eta)/2}$. Normally $\eta$ and
$\eta_u$ are extracted from the Fourier modes of height fluctuations
and in-plane phonons. Here, we demonstrate that we can use an Euler
buckling to measure these exponents directly. Current nano-fabrication
techniques can create nanoribbons as thin as $\sim2$~nm via
transmission electron microscopy~\cite{masih-ACS-10-5687-2016} and
their temperature can be controlled from as low as a few Kelvin up to
room temperature~\cite{storch-PRB-98-085408-2018}. For graphene the
thermal length at 300~K is around 3~nm, while for 1~K
$\lth\approx50$~nm.  It should therefore be possible to fabricate
ribbons with width to thermal length ratio from roughly 0.01 to 100. A
similar setup including an out-of-plane symmetry-breaking field has
been achieved experimentally~\cite{lindahl-NL-2012}. The simulations
and theory presented here provide predictions for buckling of
thermalised nanoribbons that can be tested experimentally. The
tunability of buckling via compression, temperature, and perturbing
field could be the useful for development of mechanics-based
non-volatile memories.

\section*{Declaration of competing interests}
The authors declare that they have no known competing financial
interests or personal relationships that could have appeared to
influence the work reported in this paper.
\section*{Acknowledgements}
This research was supported in part by the National Science Foundation
under grant no.  NSF-PHY-1748958. Work by PZH and DRN was also
supported through the NSF grant DMR-1608501 and via the Harvard
Materials Science Research and Engineering Center, through NSF grant
DMR-2011754.  The work of MJB was also partially supported by the
NSF through the Materials Science and
Engineering Center at UC Santa Barbara, DMR-1720256 (iSuperSeed).  DY
was supported by the Chan Zuckerberg Biohub and received funding from
the Ministerio de Econom\'ia, Industria y Competitividad (MINECO,
Spain); the Agencia Estatal de Investigaci\'on (AEI, Spain) and Fondo
Europeo de Desarrollo Regional (FEDER, EU) through grant
no. PGC2018-094684-B-C21. DY and SB thank the KITP for hospitality
during part of this project. PZH and DRN thank Abigail Plummer and
Suraj Shankar for helpful discussions. MB and DRN acknowledge helpful
conversations with Daniel Lopez. The computations in this paper were
run on the FASRC Cannon cluster supported by the FAS Division of
Science Research Computing Group at Harvard University. Some
simulations were carried out on the Syracuse University HTC Campus
Grid, which is supported by NSF-ACI-1341006.

\appendix

\section{Numerical check of the \boldmath $T=0$ theory}
\label{sec:T0}
\begin{figure}[t]
\centering
\includegraphics[width=\linewidth]{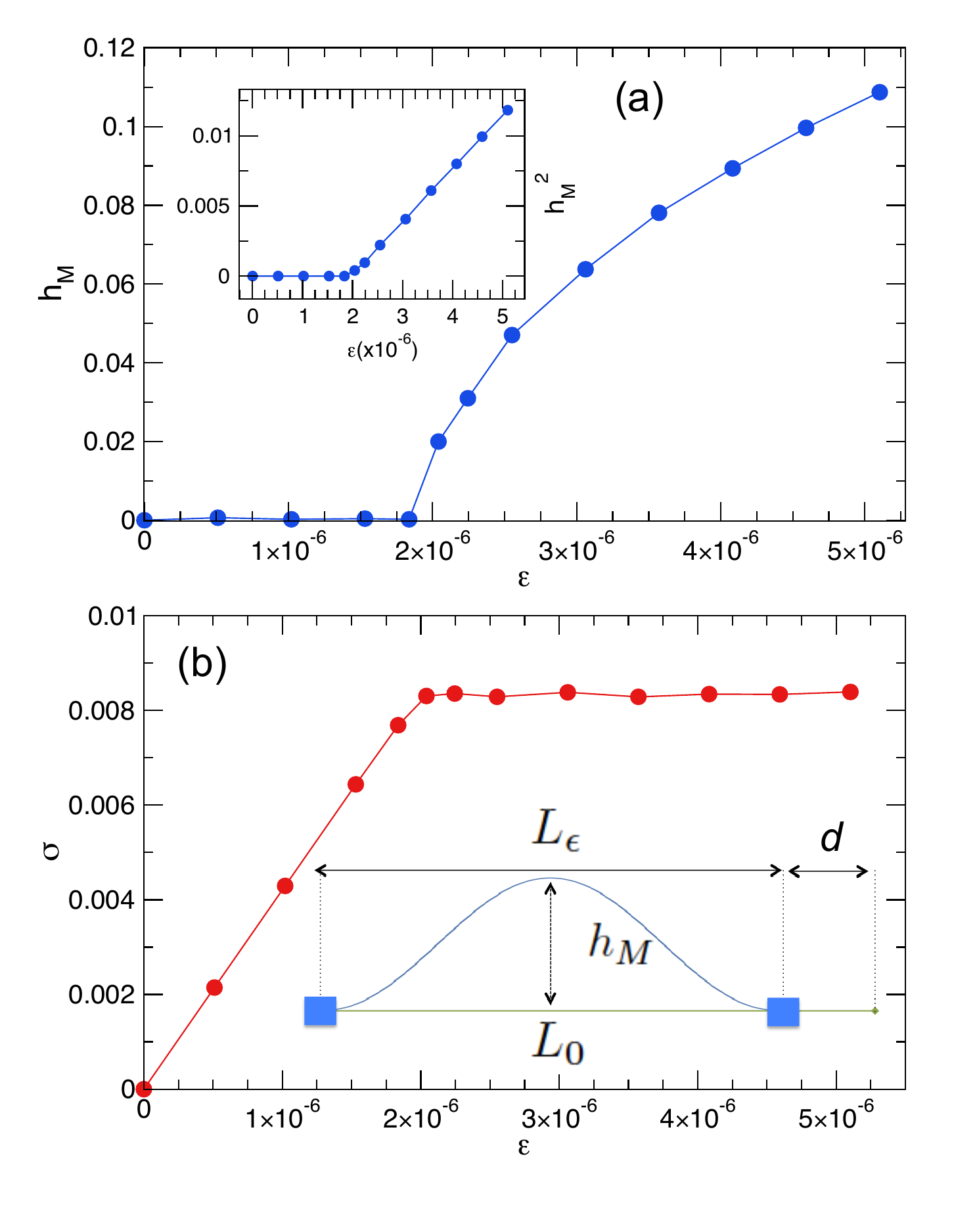}
\caption{The maximum height $h_{\rm M}$ (a) and the longitudinal
  stress $\sigma$ (b) at $T=0$ as a function of the compressive strain
  $\epsilon=1-L_\epsilon/L_0$. The inset in (a) shows the linear
  relationship between $h_{\rm M}^2$ and $\epsilon$ for
  $\epsilon>\epsilon_{\rm c}$.}
\label{fig:figSI1}
\end{figure}
To check that our coarse-grained model is consistent with the
zero-temperature theory we simulated systems with
$L_0=100a, W_0=20a, \kappasim=2.5, k/\kappasim=1440/a^2$ at $T=0$.  The energy
is minimised using the FIRE algorithm.  Recall that the connection
between continuum elastic constants and those for a triangular
lattice is $\kappabare=\frac{\sqrt{3}\hat{\kappasim}}{2}$ and
$\Ybare=\frac{2k}{\sqrt{3}}$. We plot the height amplitude $h_{\rm M}$
and stress $\sigma $ as a function of the compressive strain
($\epsilon=1-L_\epsilon/L_0$) in Fig.~\ref{fig:figSI1}. Our
simulations produce a square-root scaling of the buckling amplitude,
in agreement with the mean field theory. The computed Young's modulus, critical
stress, and critical strain are within 10\% of the theoretical
predictions
$[\frac{\Ybare^{\rm simulation}}{\Ybare^{\rm theory}}=0.99,
\frac{\sigma_\text{c}^{\text{simulation}}}{\sigma_\text{c}^{\rm theory}}=0.92,
\frac{\epsilon_\text{c}^{\text{\rm simulation}}}{\epsilon_\text{c}^\text{theory}}=0.94]$.
We attribute the small deviations to our discretised clamped
boundary conditions.

\section{Stress-strain curve}
\label{sec:stressstrain}
\begin{figure}[t]
\centering
\includegraphics[width=\linewidth]{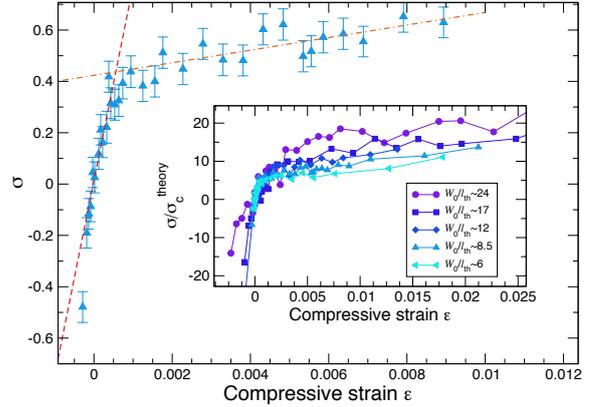}
\caption{Average stress $\sigma$ as a function of compressive strain
  $\epsilon$ for a ribbon with at a temperature large enough so that
  $W_0/\lth = 8.5$.  The dashed lines are linear fits in the
  pre-buckling (small-$\epsilon$) regime and in the post-buckling
  regime $\epsilon>\epsilon_\text{c}$. The inset shows the scaled
  stress $\sigma/\sigma_\text{c}^{\text{theory}}$ for different
  systems. The scaled critical buckling, which is proportional to the
  renormalised bending rigidity $\kappaT$, increases with increasing
  $W_0/\lth$ (or temperature). In contrast, the slope ($\YT$) becomes
  smaller with increasing $W_0/\lth$. }
\label{fig:figSI2}
\end{figure}

We fit data points close to $\epsilon=0$ to obtain $\YT$ and fit data
points beyond the buckling point to obtain the linear asymptotic
behaviour.  We use the intersection of these two lines to estimate the
critical buckling load $\sigma_\text{c}$ and critical buckling strain
$\epsilon_\text{c}$. By plotting the scaled stress
$\sigma/\sigma_\text{c}^\text{theory}$ as a function of $\epsilon$, we can see
that scaled critical buckling load
$\sigma_{\text{c}}/\sigma_{\text{c}}^{\text{theory}}$ increases with
increasing $W_0/\lth$ (increasing $T$), whereas the slope
($\YT$) decreases with increasing $W_0/\lth$, in accordance
with the theoretical expectation (see Fig.~\ref{fig:figSI2}).
\subsection{The temperature-dependent critical strain from height
  susceptibility}
\begin{figure}[t]
\centering
\includegraphics[width=\linewidth]{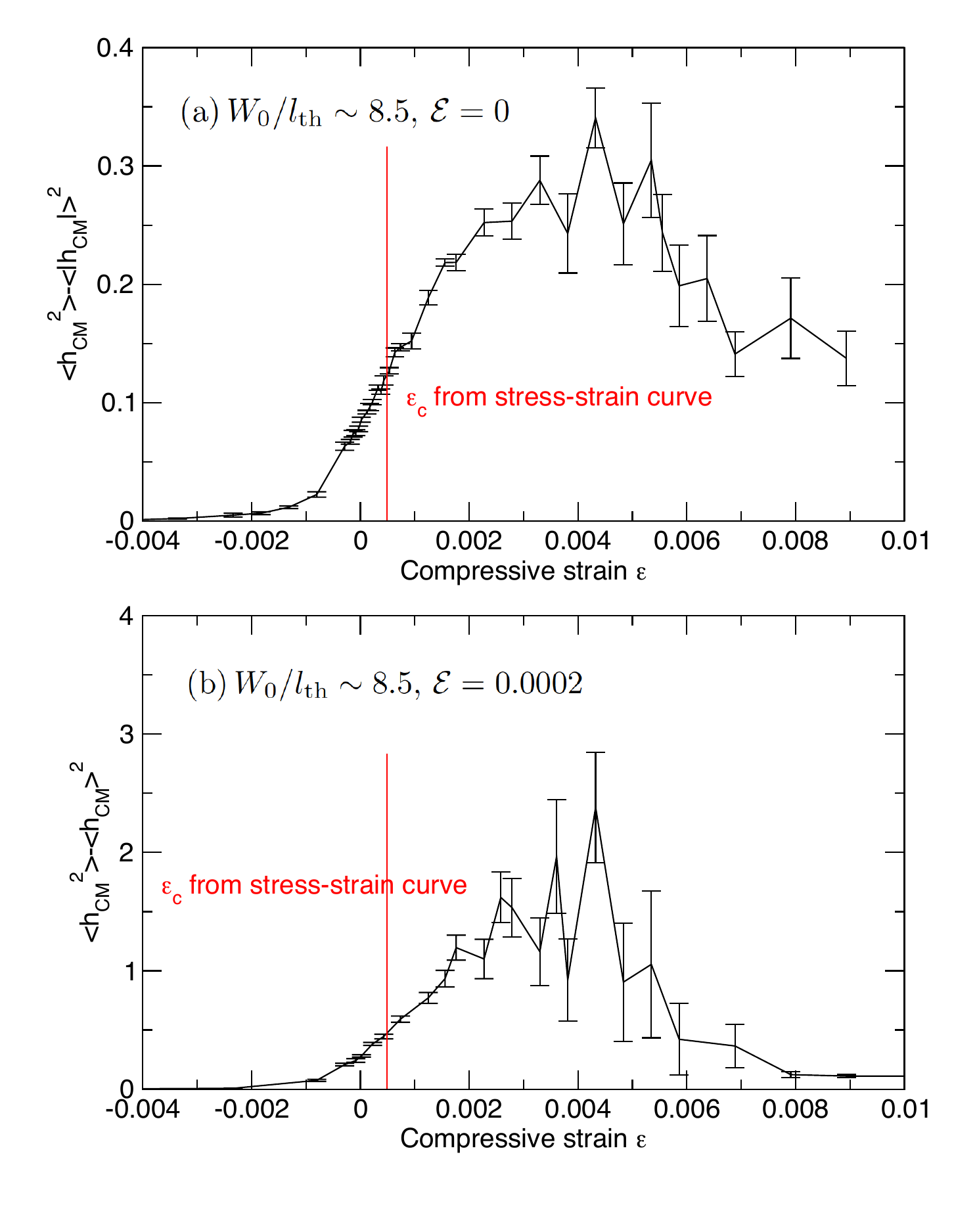}
\caption{(a) The susceptibility of the absolute value centre-of-mass
  height $\chi[|h_\text{CM}|]$ as a function of the compressive strain
  $\epsilon$ with $\mathcal{E}=0$ and (b) the susceptibility of
  centre-of-mass height $\chi[h_\text{CM}]$ as a function of the
  compressive strain $\epsilon$ with symmetry-breaking field
  $\mathcal{E}=0.0002$ for a system with $W_0/\lth\sim8.5$. The
  locations of the peaks are similar, and occur beyond the buckling
  strain $\epsilon_\text{c}$. }
  \label{fig:figSI3}
\end{figure}

Since we are interested in the buckling response due to external field
we study the height susceptibility defined as
$\chi=\dd\langle h_{\text CM}\rangle/\dd\mathcal{E}$. We can directly
obtain $\chi$ using height fluctuations with Eq.~\eqref{eq:Xi_hcm}.

As discussed in the main text, the height of center of mass
$h_\text{CM}$ beyond buckling obtained from simulations of finite
systems might flip after a long finite time. Thus $h_\text{CM}$ of
independent runs average to zero. In simulations of classical Ising
spins it is common to take the absolute value of the order
parameter~\cite{sandvik2010computational}, a strategy that can be
adopted to our problem:
\begin{equation}
\chi [|h_{\text{CM}}|]\propto \langle h^2_{\text{CM}}\rangle - \langle |h_{\text{CM}}|\rangle^2.
\label{eq:Xi_hcm_absolute}
\end{equation}
Note that this quantity differs from the \emph{true} susceptibility
(see eq~\ref{eq:Xi_hcm}). In MD simulations we can apply a small
symmetry-breaking field to bias the system to buckle in one
direction. Specifically, we simulated a system with $W_0/\lth\sim8.5$
and compare these two quantities. The function $\chi[|h_\text{CM}|]$
has a similar qualitative behavior and similar peak location to
$\chi[h_{\text{CM}}]$, as shown in fig.~\ref{fig:figSI3}. To save
computing time we use $\chi[|h_\text{CM}|]$ of
eq.~\eqref{eq:Xi_hcm_absolute} to locate the peaks.
\begin{figure}[t]
\centering
\includegraphics[width=\linewidth]{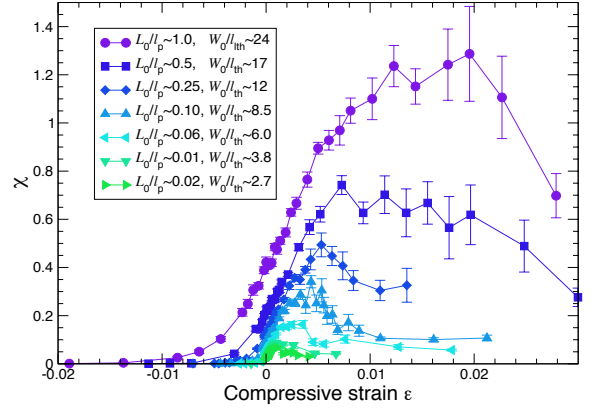}
\caption{The susceptibility of the absolute centre-of-mass 
height $\chi[|h_\text{CM}|]$ as a function of
the compressive strain $\epsilon$.
The position of the peak increases with the thermal length \lth. This is  in accordance
with the theoretical expectation of a delayed
buckling transition with increasing $T$, due to thermal
stiffening. The simulated 
systems cover the temperature range of $0.005\leq k_\text{B} T/\hat \kappa \leq 0.4$. 
\label{fig:figSI4}}
\end{figure}
The susceptibilities ($\chi[|h_\text{CM}|]$) for several temperatures
as a function of the compressive strain $\epsilon$ are plotted in
fig.~\ref{fig:figSI4}.  Here and in following plots we indicate the
temperature through the ratio of the system's width to its thermal
length, which is the appropriate scaling variable. We can clearly see
that the buckling transition persists for finite $T$, while the
position of the peaks increases with increasing $W_0/\lth$.  This
trend is consistent with our theoretical prediction that
$\epsilon_\text{c}$ should increase as the renormalisation of the
Young's modulus and bending rigidity becomes more and more
important. We find a proportionality between the critical
strains obtained from stress-strain curves and critical strains
obtained from the peaks of $\chi$; however, these critical strain
obtained from two different approaches do not coincide exactly (see
Fig.~\ref{fig:figSI5}).
\begin{figure}[t]
\centering
\includegraphics[width=\linewidth]{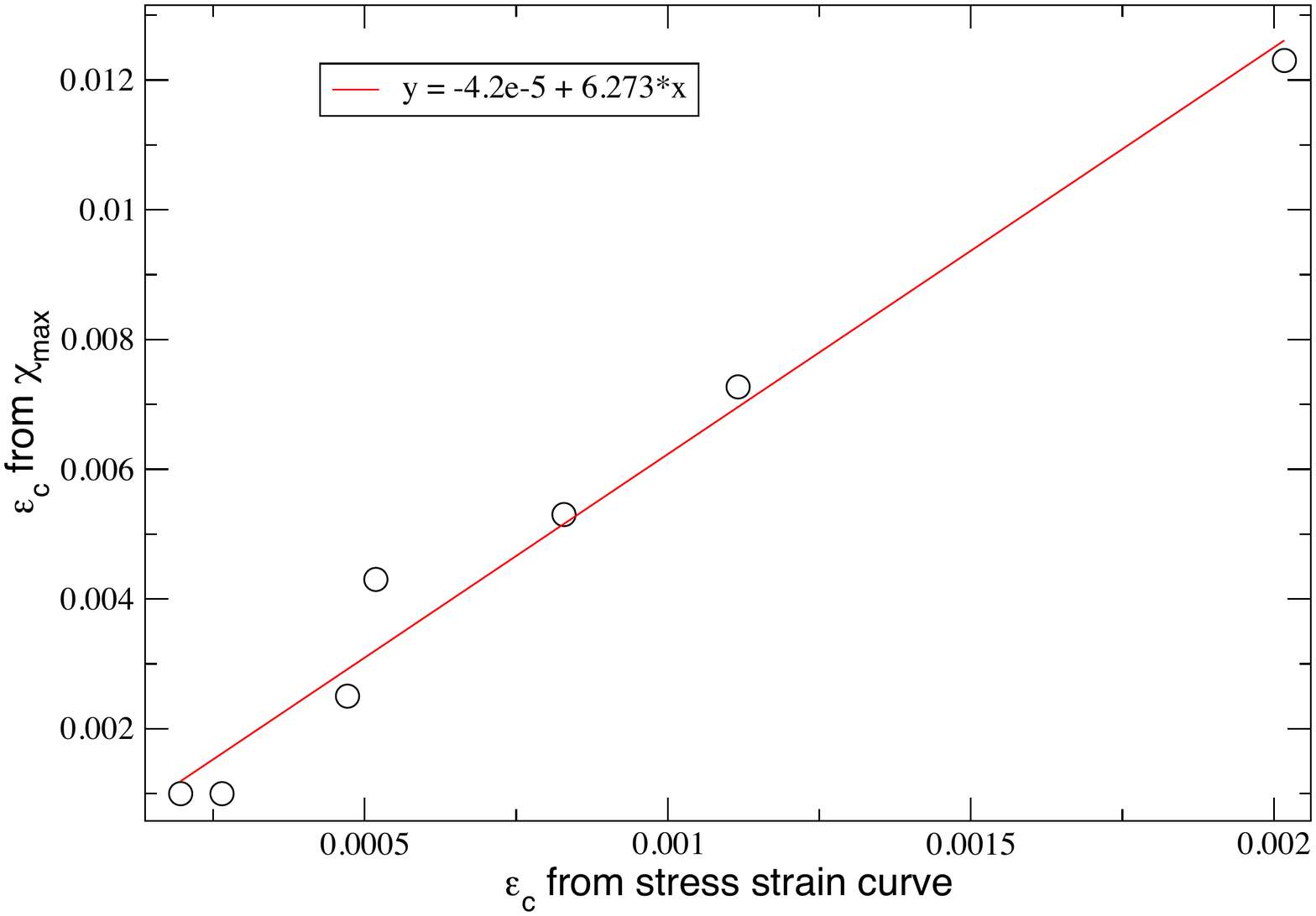}
\caption{Critical strains obtained from the peaks of $\chi$
  (eq.~\eqref{eq:Xi_hcm_absolute}) as a function of critical strains
  obtained from stress-strain curves. }.
\label{fig:figSI5}
\end{figure}

\section{Variational approach}
\label{sec:variational}
Here we describe how eliminating in-plane displacement fields leads to a non-local stretching term in the Gibbs
energy. For a clamped 1D ribbon we write the effective 1D Young's
modulus as $Y_{\rm 1D}=\Ybare W_0$ and the bending rigidity as
$\kappa_{\rm 1D}=\kappabare W_0$.  The amount of work is $-Fd$ and the
compression distance $d$ can be approximated as follows, 
\begin{align}
L_{\epsilon}+d&\simeq\int_{-L_\epsilon/2}^{L_\epsilon/2}\sqrt{1+\left(\frac{\dd h}{\dd x}\right)^2}\dd x\\
d&\simeq\frac{1}{2}\int_{-L_\epsilon/2}^{L_\epsilon/2} \left(\frac{\dd h}{\dd x}\right)^2 \dd x.
\end{align}
We assume variations only in the $x$-direction. The Gibbs free energy
is given by
\begin{equation}
\begin{split}
 G[u_x, h]=&\frac{\kappa_{\rm 1D}}{2}\int_{-L_\epsilon/2}^{L_\epsilon/2} \dd x\left(\frac{\dd^2h}{\dd x^2}\right)^2\\
&+\frac{Y_{\rm 1D}}{2}\int_{-L_\epsilon/2}^{L_\epsilon/2}\dd x\left[\frac{\dd u}{\dd x}+\frac{1}{2}\left(\frac{\dd h}{\dd x}\right)^2\right]^2\\
&-\frac{F}{2}\int_{-L\epsilon/2}^{L\epsilon/2} \dd x\left(\frac{\dd h}{\dd x}\right)^2,
\end{split}
\label{eq:G0}
\end{equation}
where $F=W_0\sigma_{xx}$.  We will now focus on the middle stretching
term $G_{\rm s}$ controlled by $Y_{\rm 1D}$. As is typically done in the 2D case,
we focus on the vector-potential-like contribution,
$A(x)=\frac{1}{2}\left(\frac{\dd h}{\dd x}\right)^2$, and we write the
fields in Fourier space as,
\begin{align}
\frac{\dd u}{\dd x}=U_0 + \sum_{q\neq 0}\ii q\tilde{u}(q)\ee^{\ii qx}\\
A=A_0 + \sum_{q\neq 0}\tilde{A}(q)\ee^{\ii qx},
\end{align}
\label{eq:Fourier}
where we have separated out the $q=0$ modes. The stretching energy $G_{\rm s}$ is given by 
\begin{align}
G_{\rm s}=&\frac{Y_{\rm 1D}}{2}\int_{-L_\epsilon/2}^{L_\epsilon/2}\dd x \left[U_0 + A_0+ \sum_{q\neq 0}\left(\ii q\tilde{u}(q)  +\tilde{A}(q) \right)\ee^{\ii qx}\right]\nonumber \\
&\quad \times
 \left[U_0 + A_0+ \sum_{q\neq 0}\left(\ii q'\tilde{u}(q')  +\tilde{A}(q') \right) \ee^{\ii q'x}\right]\nonumber\\
=&\frac{Y_{\rm 1D}}{2}L_\epsilon(U_0+A_0)^2\nonumber \\
&\quad +\frac{Y_{\rm 1D}}{2}\sum_{q\neq 0}\sum_{q'\neq 0}\int_{-L_\epsilon/2}^{L_\epsilon/2} \ee^{\ii(q+q')x}\left(\ii q\tilde{u}(q)  +\tilde{A}(q) \right)\nonumber \\
&\qquad\qquad\qquad\qquad\qquad\times\left(\ii q'\tilde{u}(q')  +\tilde{A}(q') \right)\dd x \nonumber\\
=&\frac{Y_{\rm 1D}}{2}L_\epsilon(U_0+A_0)^2+\frac{Y_{\rm 1D}L_\epsilon}{2}\sum_{q\neq 0}|\ii q\tilde{u}(q)  +\tilde{A}(q)|^2.
\label{eq:Gss}
\end{align}
The stretching energy $G_s$ is clearly minimised when $\tilde{u}(q)=-\ii \tilde{A}(q)/q$.
Upon imposing constant strain and  the boundary conditions we find, 
\begin{align}
U_0=&\frac{1}{L_\epsilon}\int_{-L_\epsilon/2}^{L_\epsilon/2}\dd x\ \frac{\dd u}{\dd x}
=\frac{1}{L_\epsilon}[u(L_\epsilon/2)-u(-L_\epsilon/2)]
=0,\\
\intertext{and similarly,}
A_0=&\frac{1}{L_\epsilon}\int_{-L_\epsilon/2}^{L_\epsilon/2}\dd x\ A
=\frac{1}{L_\epsilon}\int_{-L_\epsilon/2}^{L_\epsilon/2}\frac{1}{2}\left(\frac{\dd h}{\dd x}\right)^2 \dd x.
\end{align}
%\end{widetext}
Upon substituting Eq.~\eqref{eq:Gss} in the form
$G_{\rm s}=\frac{\Ybare_{\rm 1D}}{2}L_\epsilon A_0^2$ into
Eq~\ref{eq:G0}, we obtain the free energy of Eq (10) of the main text,
provided we include a contribution from the symmetry-breaking field.

We now discuss an out-of-plane field that couples to the height. For instance,
if we put uniform charges on the ribbon and place it within a
uniform electric field, the potential energy is
$V_{\perp}=-\int_{-L_\epsilon/2}^{L_\epsilon/2}\rho\mathcal{E}\,h\,\dd x$. After
collecting terms, including an out-of-plane external electric field
$\mathcal{E}$, we obtain the Gibbs energy
%\begin{widetext}
\begin{equation}
\begin{split}
 G[h, \mathcal{E}]=&\frac{\kappa_{\rm 1D}}{2}\int_{-L_\epsilon/2}^{L_\epsilon/2}\dd x\left(\frac{\dd^2h}{\dd x^2}\right)^2\\
& +\frac{Y_{\rm 1D}}{2L_\epsilon}\left[\int_{-L_\epsilon/2}^{L_\epsilon/2}\dd x\frac{1}{2}\left(\frac{\dd h}{\dd x}\right)^2\right]^2\\
& -\frac{F}{2}\int_{-L_\epsilon/2}^{L_\epsilon/2}\dd x\left(\frac{\dd h}{\dd x}\right)^2-\rho\mathcal{E}\int_{-L_\epsilon/2}^{L_\epsilon/2} h\ \dd x.
\end{split}
\label{eq:Gh}
\end{equation}
%\end{widetext}
Close to the buckling transition we focus on the first buckling mode
$h=h_{\rm M}\frac{1}{2}\left[1+\cos \left(\frac{2\pi
      x}{L_\epsilon}\right)\right]$,
where $h_M$ is the height amplitude, as an ansatz that insures
tangential boundary conditions
$\left. \frac{\dd h(x)}{\dd x}\right|_{x\pm=L_{\epsilon}/2}=0$. We
then obtain Eq.~(11) of the main text,
 %\small
%\begin{widetext}
\begin{equation}
  G=\frac{\pi^2}{4L_\epsilon}\left(\frac{4\kappa_{\rm 1D}\pi^2}{L_\epsilon^2}-F\right)h^2_{\rm M}+\frac{\pi^4Y_{\rm 1D}}{32L_\epsilon^3}h^4_{\rm M}-\frac{\rho L_\epsilon \mathcal{E}}{2}h_{\rm M}.
%\normalsize
\end{equation}
It is helpful to write the above equation in terms of new parameters $a, b, \epsilon_\text{c}$
\begin{equation}
  G=a(\epsilon-\epsilon_\text{c})h^2_{\rm M}+bh^4_{\rm M}-\frac{\rho L_{\epsilon} \mathcal{E}}{2}h_{\rm M}, 
\end{equation}
where $a=\frac{Y_{\rm 1D}\pi^2}{4L_{\epsilon}}$,
$b=\frac{\pi^4Y_{\rm 1D}}{32L_{\epsilon}^3}$,
$\epsilon_\text{c}=\frac{4\pi^2\kappa_{\rm 1D}}{Y_{\rm
    1D}L_\epsilon^2}$.
Upon minimising the Gibbs free energy by setting $\left. \frac{\dd G}{\dd h_M}\right|_{\mathcal{E}=0}=0$, we find
%\begin{widetext}
\begin{align}
h_M &=
  \begin{cases}
    0,    & \text{if } \epsilon<\epsilon_{\rm c}\\
    \pm\sqrt{\frac{a}{2b}(\epsilon-\epsilon_{\rm c}})=\frac{2L_{\epsilon_\text{c}}}{\pi}\sqrt{\epsilon-\frac{4\kappabare\pi^2}{\Ybare L^2_{\epsilon_\text{c}}}},& \text{if } \epsilon>\epsilon_{\rm c}
  \end{cases}
\end{align}
%\end{widetext}

\subsection{Susceptibility}
To obtain the susceptibility at zero external field we first solve
$\frac{\dd G}{\dd h_{\rm M}}=0$, which leads to
\begin{align}
0=&2ah_M(\epsilon_\text{c}-\epsilon)+4bh^3_{\rm M}-\frac{\rho L_{\epsilon} \mathcal{E}}{2}\\
\mathcal{E}=&\frac{2}{\rho L_\epsilon}[4bh^3_{\rm M}+2a(\epsilon_\text{c}-\epsilon)h_{\rm M}].
\end{align}
We can now calculate the susceptibility and use $\rho=Q/L_0$ and $Y_{\rm 1D}=\Ybare W_0$ to obtain
%\begin{widetext}
\begin{align}
\left. \frac{\partial h_{\rm M}}{\partial \mathcal{E}}\right|_{\mathcal{E}=0} &=
  \begin{cases}
    \frac{Q}{\Ybare\pi^2}\frac{L_{\epsilon_\text{c}}}{W_0}(\epsilon_{\rm c}-\epsilon)^{-1}    & \quad \text{if } \epsilon<\epsilon_{\rm c}\\
    \frac{Q}{2\Ybare\pi^2}\frac{L_{\epsilon_\text{c}}}{W_0}(\epsilon-\epsilon_{\rm c})^{-1} & \quad \text{if } \epsilon>\epsilon_{\rm c}
  \end{cases}
\end{align}
%\end{widetext}
For a ribbon in a gravitational field simply replace $\mathcal{E}=g$ and $Q=m$, where $m$ is the total mass.

\end{document}